\def\year{2020}\relax
\documentclass[letterpaper]{article} 
\usepackage{aaai20}  
\usepackage{times}  
\usepackage{helvet} 
\usepackage{courier}  
\usepackage[hyphens]{url}  
\usepackage{graphicx} 
\usepackage{graphicx}  
\usepackage{booktabs}
\usepackage{array}
\newcolumntype{L}{>{\arraybackslash}m{7.6cm}}
\title{Analysing the Extent of Misinformation in Cancer Related Tweets}
\author{Rakesh Bal\textsuperscript{\rm 1}\thanks{Equal contribution}, Sayan Sinha\textsuperscript{\rm 1$*$}, Swastika Dutta\textsuperscript{\rm 1}, Rishabh Joshi\textsuperscript{\rm 2}, Sayan Ghosh\textsuperscript{\rm 3}, Ritam Dutt\textsuperscript{\rm 2}\\ \Large \\ 
\textsuperscript{\rm 1}Indian Institute of Technology Kharagpur, \textsuperscript{\rm 2}Carnegie Mellon University, \textsuperscript{\rm 3}University of North Carolina at Chapel Hill\\
\textsuperscript{\rm 1} \{rakesh.bal, sayan.sinha, swastika\}@iitkgp.ac.in,\\ \textsuperscript{\rm 2} \{rjoshi2, rdutt\}@andrew.cmu.edu,\\  \textsuperscript{\rm 3} sayghosh@cs.unc.edu \\
}
 
\begin{document}

\maketitle
\begin{abstract}
Twitter has become one of the most sought after places to discuss a wide variety of topics, including medically relevant issues such as cancer. This helps spread awareness regarding the various causes, cures and prevention methods of cancer. However, no proper analysis has been performed, which discusses the validity of such claims. In this work, we aim to tackle the misinformation spread in such platforms. We collect and present a dataset regarding tweets which talk specifically about cancer and propose an attention-based deep learning model for automated detection of misinformation along with its spread. We then do a comparative analysis of the linguistic variation in the text corresponding to misinformation and truth. This analysis helps us gather relevant insights on various social aspects related to misinformed tweets.
\end{abstract}
\section{Introduction}
Social networks play a vital role in information dissemination among a large mass of people. These platforms are easily accessible and can be used to express views and information comfortably. Various microblogging sites like Twitter allow people to share their opinion on topics ranging from personal issues to major political events such as elections. Healthcare is one such popular domain where people freely share their thoughts and experiences. However, on these platforms, there is no check to ensure the sanctity of the content posted, and thus the detection and control of misinformation is a task of utmost importance. To this end, we present a new dataset consisting of tweets about various aspects of cancer treatment and causes. We also formulate deep learning models to detect such tweets automatically. Our contributions to this paper can be summarized as follows:
\begin{itemize}
\item We propose a new dataset to identify prevalent causes and cures related to cancer in online social media (tweets). 
\item We develop algorithms for identifying ``medically relevant" tweets and propose an attention-based sequence labelling model for extracting objects said to cause and cure cancer.
\item We apply our proposed algorithms on our entire Twitter corpus to identify terms frequently associated with cancer. We also analyse the linguistic differences in the text of informed and misinformed tweets. Our analysis provides an insight into the social aspects related to the spread of misinformation related to cancer.
\end{itemize}

\section{Related Work}
Misinformation spread in social media is a common nuisance in modern times, and this has affected our beliefs and social constructs ranging from politics to healthcare. \cite{allcott2019trends} discusses the role of misinformation diffusion in politics and how it has affected democratic institutions. The authors in this work use two datasets of social media posts from Facebook and Twitter and discuss the trend in the diffusion of fake content in these social platforms. On a similar note \cite{shin2018diffusion} also discuss the role of misinformation discussion in the political scenario and how old rumours are again resurfaced as ``news". \cite{anspach2018believe} discuss in their work of how the people are misinformed about various news or political scenarios by only reading bits off social media posts.  \cite{bode2018see} shows how false news gets spread about global health topics in social media. The authors study ways to limit the spread of misinformation spread by using various correction measures and doing a human study to understand how effective a correction algorithm is in doing so. \cite{ghenai2018fake} in their work build a classifier to identify the users who are prone to spread false information in Twitter about treatment techniques for cancer by using several user based attributes. 

There have been several studies on analysing misinformation on YouTube videos across different medical domains like prostate cancer \cite{loeb2019dissemination}, orthodontics \cite{kilincc2019assessment}, and idiopathic pulmonary fibrosis \cite{goobie2019youtube}. The studies have highlighted that despite having colossal user-engagement, popular YouTube videos on medical issues are plagued with biases, and inaccuracies and generate unreliable and poor-quality information. 

Likewise, there have been seminal work in understanding the multi-faceted nature of health misinformation such as temporal trends in anti-vaccine discourse \cite{gunaratne2019temporal}, identifying people susceptible to spread misinformation \cite{ghenai2018fake}, the trajectories of information spread of Ebola in Twitter \cite{liang2019did}, and the bursty nature of rumour related topics \cite{ghenaicatching}. 
In this work, we look at the problem from an information extraction perspective wherein we attempt to identify medically relevant tweets pertaining to cancer from the Twitter stream and automate the extraction of causes and treatments related to cancer. 



\section{Dataset Description}
We used the Twitter Streaming API\footnote{https://developer.twitter.com/en/docs}, to collect tweets having certain \textbf{keywords} in conjunction with cancer. These keywords were manually identified from the work of \cite{difonzo-cancer}, which recognises common words associated with cancer rumours on online social media boards. The initial set of keywords included ``cure", ``family", ``attack", ``cause", ``radiation" etc., to conform with the different set of rumour categories as mentioned in \cite{difonzo-cancer}. The tweets were collected for 29 days, from 11th June 2018 to 9th July 2018.

After removing retweets and duplicates, we finally obtained 20,137 unique tweets. Out of these, $35.87\%$ mentioned about objects which cause cancer, $53.32\%$ talked about entities which cure cancer, whereas the rest involved discussions on things which help prevent cancer.
For this particular study, we decided to use tweets which mention `cause' and `cure' related to cancer since they occur more frequently. We deliberately chose to omit tweets which talk about the prevention of cancer due to reasons as discussed under \textit{Preliminary Analysis}\ref{label:preliminary_analysis}.


\subsection{Preprocessing}
We perform standard preprocessing tasks on the tweets before using them for further analysis. We use regular expressions to remove URLs, mentions, and emoticons. We also segment hashtags based on Camel Case\footnote{\url{https://en.wikipedia.org/wiki/Camel_case}} and common heuristics applied in similar works since hashtags often contain topical information related to a tweet.

\subsection{Identifying medically relevant tweets} 
The initial filtering of cancer-related tweets was accomplished via specific keywords. Thus, there were several tweets which lacked any relevant medical information related to cancer. We define a tweet to be \textit{``medically relevant"} if it mentions any object as a cause or cure for cancer. Consequently, we define a tweet to be \textit{``medically non-relevant"} if it fails to mention any object which causes cancer, cures cancer or uses cancer as a metaphor for some other scenario.  We illustrate some ``medically relevant" tweets and ``medically non-relevant" tweets in Table 1.
We leverage human annotation to distinguish between the medically relevant and non-relevant tweets on the part of the dataset. Three annotators were employed, each of whom is competent in English and use Twitter regularly, but none of whom is the author of this paper. The results were determined using majority voting. We identify 408, 207 and 386 tweets as medically relevant for cause, cure and prevent respectively. Similarly, we identify 590, 1297 and 184 tweets as non-medically relevant for cause, cure and prevent respectively. We automate the process of classifying such medically relevant tweets and describe it further in the next section.

\begin{table*}[htbp]
\caption{Examples of medically relevant and non-relevant tweets}
\centering
\scriptsize
\begin{tabular}{c L L}
\toprule
Domain & Medically Relevant & Medically Non-relevant \\ 
\midrule

Causes & 3 weeks ago, two papers, and the news reports that followed, caused a short-lived scare that \#CRISPR causes cancer. This wasn't the first CRISPR scare, and it won't be the last. & Cancer is my biggest fear. Cause you could live the most moral life, exercise, eat well and still one day you wake up and your life is a mess. HIV is a choice (in most cases). \\ \hline
Cures & The major demand for rhino horn is in Asia where it's used in ornamental carvings and traditional medicine. Rhino horn is touted as a cure for hangovers, cancer and impotence. Science has proven all this to be false. & No matter what our president does, he will never get the credit he deserves. He could cure cancer, and everyone would say.. he should of been worrying about heart disease.   \#TrumpKimSummit \\ \hline
Prevents & Garlic has sulfur compounds such as allicin which may help prevent infections by blocking specific enzymes. There is research that links garlic intake to a decreased risk of cancer, specifically stomach, colon and esophagus. \#diet \#nutrition \#food \#Health  & Uganda Cancer Institute and Childcare for Cancer Foundation start a drive to equip students across the country with knowledge to prevent and fight cancer. This and more on \#NTVATONE coming up \\ 


\bottomrule
\end{tabular}
\label{}
\end{table*}

\subsection{Extracting the anchors}
Having identified the ``medically relevant" tweets, we ask the same set of annotators to extract the objects which are mentioned to cause or cure cancer respectively, on the same portion of the dataset. These objects are hereby referred to as ``anchors".
We also discuss automating the process of identifying these anchors in the subsequent section. 
\begin{figure}[htb]       
    \centering
    \includegraphics[width=\linewidth]{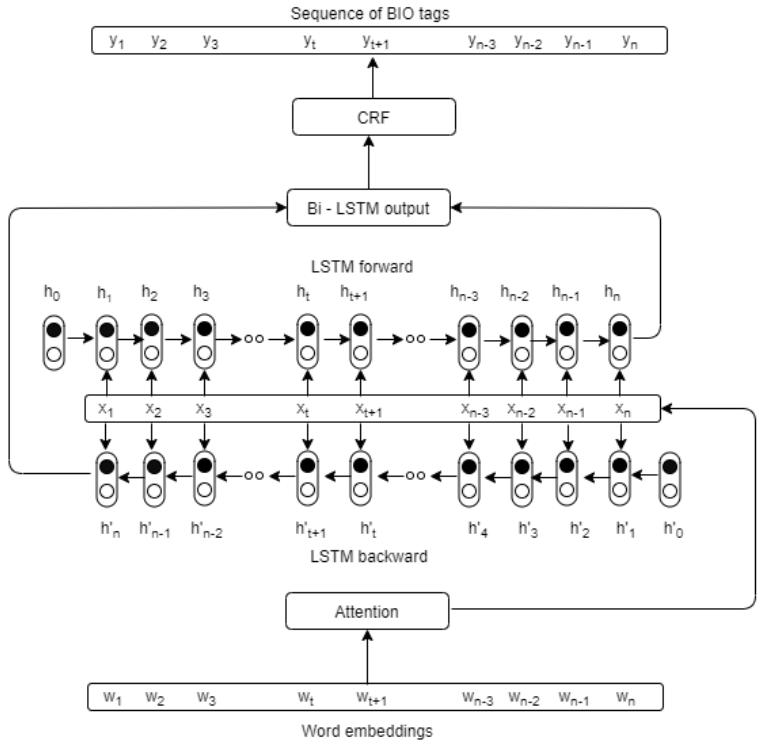}
\vspace{-4mm}
\caption{Deep Learning model used for the BIO tagging. }
    \label{fig:icwsm_screenshot}
\vspace{-6mm}
\end{figure}

\subsection{Preliminary analysis}
\label{label:preliminary_analysis}
On performing an initial analysis on the partly annotated dataset, we discover certain approximate percentages of the tweets which were spreading misinformation. This is done by manually checking the anchors of $1000$ randomly selected tweets and verifying if they are indeed causing, curing or preventing cancer. The results show that approximately $28\%$ of the tweets related to causes of cancer, $55\%$ among the ones which discuss curing cancer and $5\%$ of the tweets which discuss preventing cancer can be classified as misinformation. Thus, we choose to focus our research on tweets which talk about causing or curing cancer only. We neglect ``prevents" owing to the fact that the initially determined approximate percentage of misinformation is pretty less. This result helps us get an insight that any activity which has a positive effect on the mind and body could help in preventing cancer. We also find that across all the three categories, the ratio of positive to negative stance was of the order of $50:1$. This allowed us to specifically exclude stance detection since their presence was by and large consistent.

\section{Methodology}
\subsection{Classifying medically relevant and non-relevant tweets}
We handle this classification task by using a simple feed-forward fully connected deep learning network with $3$ hidden layers -- each having $1024$, $512$ and $256$ neurons respectively using binary cross-entropy loss. For analysis, we experimented with both tfidf and tfidf weighted PubMed \cite{zhang2019biowordvec} \cite{chen2019biosentvec} pretrained embeddings to convert the tweets to sentence vectors. For our objective, the latter was determined to be more suitable. 
These sentence vectors are then fed into our model and trained to output the label of the tweet, i.e., $1$ for medically relevant and $0$ for medically irrelevant tweets. The dataset was divided into $4:1$ ratio for training and validation.
On running the model on the entire dataset, we discover that around $42\%$ of the tweets were medically relevant.
\subsection{Attention based sequence labelling to detect anchors.}
Sequence labelling model \cite{graves2012supervised} is a generative framework used to map an arbitrary length sequence of inputs to corresponding labels.
In this case, we learn the context of the tweet from the sequence of words in it and then generate the list of BIO (Begin-Inside-Outside) tags, as discussed in the next subsections. The BIO tags help to demarcate the ``anchors" in the tweet. For example, \\

\fbox{
\parbox{0.9 \linewidth}{
\textbf{Tweet:} ``Processed meats causes cancer according to \#WHO."
\\
\textbf{BIO Tags:} \{B-anchor, I-anchor, O, O, O, O, O\}
}
}
\subsection{Automating detection of tweets discussing causes of cancer}
In this work, we propose an attention-based BiLSTM-CRF model \cite{huang2015bidirectional} for the detection of anchor words (Figure \ref{fig:icwsm_screenshot}). We use pretrained PubMed word embeddings for each of the words in a tweet, weight them through an attention network, and feed them through a Bidirectional LSTM. Finally, Conditional Random Field (CRF) \cite{lafferty2001conditional} is used before producing the output. We used a simple attention network featuring two hidden layers with $64$ and $32$ neurons each.

We also try out other model variants such as BiLSTM with softmax, simple CRF, etc. We experimented with self-attention inspired from transformer \cite{vaswani2017attention} encoder network. We also attempted various kinds of word embeddings, such as PubMed and skip-gram embedding, learnt from the text itself. The results have been presented in Table 3. As evident, the results from attention-based BiLSTM-CRF with PubMed embeddings provided the best results.


\begin{table}[htbp]
\vspace{-6mm}
\caption{Medical relevance detection}
\centering
\scriptsize
\scalebox{1}{
\begin{tabular}{c|c c|c c}
\toprule
& \multicolumn{2}{c|}{\textbf{Simple tfidf}} & \multicolumn{2}{c}{\textbf{tfidf weighted PubMed}}\\ 
\midrule
{\textbf{Domain}} & {\textbf{F1 score}} & \textbf{Accuracy} & {\textbf{F1 score}} &\textbf{Accuracy}\\ \hline
causes & $0.7238$ & $0.8350$ & $\textbf{0.7818}$ & $0.8400$\\ 
prevents & $0.8145$ & $0.7310$ & $\textbf{0.8341}$ &$0.7719$\\ 
cures & $0.4117$ & $0.9117$ & $\textbf{0.5106}$ & $0.8985$\\ 
\bottomrule

\end{tabular}}
\label{med_rel_result}
\vspace{-6mm}
\end{table}

\subsection{Automating detection of tweets discussing cures of cancer}

The above-mentioned methodologies were also implemented for tweets which discuss about curing cancer. However, we found that the results from these approaches were not up to the mark. Hence, we took an alternate approach.

According to the National Cancer Institute \cite{Treatmen46:online}, \textit{chemotherapy}, \textit{radiation therapy}, \textit{immunotherapy}, \textit{targeted therapy}, or \textit{hormone therapy} are the very specific techniques that are recognised to potentially cure cancer. Misinformation can be detected in tweets referring to ``cures" of cancer by determining the presence of objects or techniques similar to this fixed set of proven cures by comparing their similarity in a vector space.
This enabled us to formulate our approach towards detecting ``anchors" in tweets which talk about curing cancer. The word vectors, in their PubMed embedding vector form, are compared with the word vectors of the exhaustive list of items that cure cancer. We take two simple linear combinations of the dimensions of the vector space spanned by the word embeddings. If the first linear combination is higher than the second, we conclude the presence of a word in the tweet, which can cure cancer. 
Absence of any such ``similar" word signifies that the tweet speaks about a technique or item that is not proven to cure cancer.

\begin{table}[htbp]

\caption{Results from various methods on anchor detection for causes and cures of cancer}
\centering
\scriptsize
\scalebox{0.95}{
\begin{tabular}{c c c c c}
\toprule
{\textbf{Method}}  & \textbf{F1 score - Causes} &  \textbf{F1 score - Cures}\\ 
\midrule
CRF (skip-gram) & $0.5917$ & $0.5205$\\ 
CRF (PubMed)  & $0.4671$ & $0.3712$\\ 
CRF (skip-gram + POS)  & $0.5936$ & $0.5170$\\ 
CRF (skip-gram + POS + deptag)  & $0.6406$  & $0.5467$\\ 
CRF (PubMed + POS + deptag) & $0.4841$ & $0.3774$\\ 
BiLSTM-Softmax (skip-gram) & $0.5101$ & $0.5586$\\ 
BiLSTM-CRF (PubMed + POS) & $0.6321$ & $0.5673$ \\ 
BiLSTM-CRF (PubMed + POS + deptag) & $0.6421$ & $0.4662$\\ 
BiLSTM-CRF (skip-gram) & $0.6017$ & $0.4541$\\ 
BiLSTM-CRF (PubMed) & $0.6667$ & $0.4968$\\ 
\textbf{Simple attention BiLSTM-CRF (PubMed)} & $\textbf{0.6846}$ & $\textbf{0.5821}$ \\ 
Self attention BiLSTM-CRF (PubMed) & $0.6573$ & $0.5520$\\ 

\bottomrule
\end{tabular}}
\label{attn_result}
\end{table}

\section{Results}
\subsection{Detection of medically relevant tweets}
As evident from the Table 2 using tfidf weighted PubMed outperformed using simple tfidf for word representation. We obtain F1 score of $0.7818$ for ``causes", $0.8341$ for ``prevents", and $0.5106$ for ``cures" using tfidf-weighted PubMed embeddings.

\subsection{Analysing tweets discussing causes of cancer}
We curated the top 20 most discussed keywords on things which cause cancer. We discovered that the keywords which account for misinformation comprise $38.05\%$ of all the tweets concerned. The approach of considering keywords as the basis of estimating misinformation highlights its \textit{spread} as we have taken into consideration only those topics which are most discussed on. This is significantly greater than the previous analysis of $28\%$ (subsection - \textit{Preliminary Analysis}), which was on the basis of the number of tweets containing misinformation. Thus, we can see that the spread of misinformation is higher compared to the number of misinformed tweets. Top keywords featuring in misinformed tweets were detected as ``sun", ``meat", ``coffee", ``bacon", ``sunscreen", ``crispr", ``sugar" etc. As mentioned earlier, the best F1 score of $0.6846$ was achieved using an attention-based BiLSRM CRF model.

\subsection{Analysing tweets discussing cures of cancer}
We also curated the top 20 most discussed keywords on things which cure cancer. In this case, We discovered that the keywords which account for misinformation comprise $76.19\%$ of all the tweets concerned. Once again, the spread of misinformation based on keyword analysis is significantly greater than the previous estimate of $55\%$ (subsection - \textit{Preliminary Analysis}). Top keywords featuring in misinformed tweets were detected as ``cannabi", ``hemp", ``dog 's urin", ``carrot juic", ``nimbolid", ``herbal", ``immunotherapy" etc. As mentioned earlier in Table 3, the best F1 score of $0.5821$ was achieved using an attention-based BiLSRM CRF model. However, we took an alternate approach as described in the subsection \textit{Automating detection of tweets discussing cures of cancer} and found the F1 score to be $0.7363$.


\section{Discussion}

\begin{figure}[htb]       
    \centering
    \includegraphics[width= \linewidth]{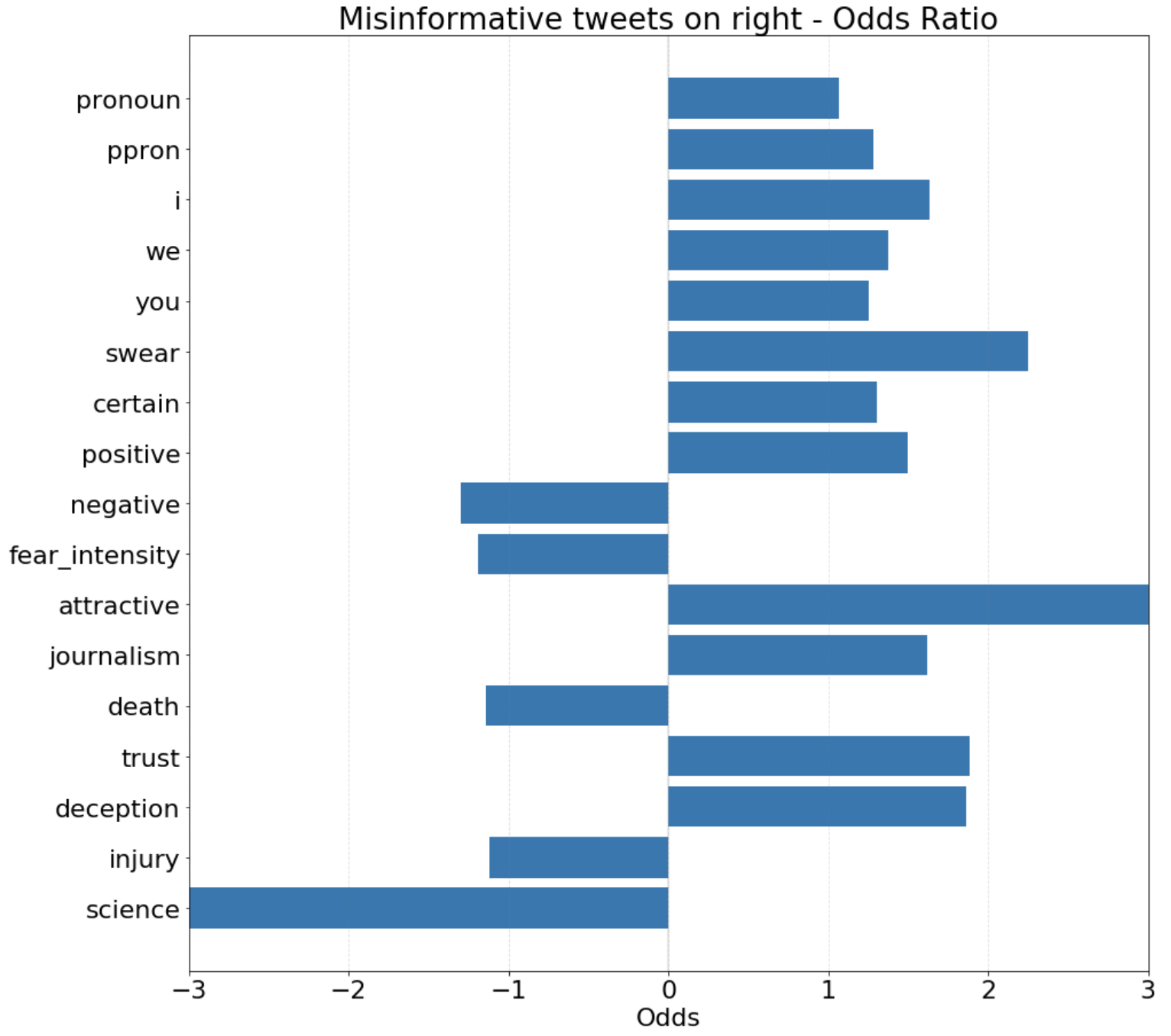}
\caption{Linguistic variations between correct tweets (left) and the tweets with misinformation (right). }
    \label{fig:oddsratio}
\vspace{-3mm}
\end{figure}

We derived certain insights based on the social aspects. We separated tweets which were spreading misinformation from the ones with correct information and tried to see the linguistic variations between the two groups. We performed simple statistical T-tests to find statistically significant (two-tailed p-values $ < 0.05$) variations between the two groups. We used LIWC \cite{liwcpaper}, NRC lexicons based on Word emotion \cite{wordemotionpaper} and Valence, Arousal and Dominance \cite{nrcvadpaper}. We also used Empath \cite{empathpaper} and extracted a total of 251 different features from the texts. Here, we mostly discuss a few social terminologies which were recognised in this process and also present the odds ratio for each of them (given in brackets). The complete odds ratio is given in Figure \ref{fig:oddsratio}.

First, we find that tweets containing misinformation were \textit{attractive} in nature ($3.770$). Also, they are prone to the notice of \textit{journalists} ($1.616$). These could be potential reasons as to why the spread of such tweets have drastically increased compared to their initial count. Text backed by news agencies, though it might contain misinformation, is likely to gain more attention. Moreover, with the advancement of scientific prowess, certain accepted attributes related to cancer might change over time. Hence a tweet, which had provided correct information at an earlier time, could be misleading in the future. One such example has been presented in the first tweet in Table 1, under the \textit{Medically Relevant} column.

\textit{Trust} issues ($1.884$) and \textit{deception} ($1.860$) also regularly featured in tweets containing misinformation. Apparently, both these attributes are interlinked, given that misleading information would naturally be pretentious.

Certain other aspects, which we found to be common in misinformed tweets were \textit{swear terms} ($1.920$) and \textit{joy} ($1.287$). On the flip side, \textit{sadness} ($-1.103$) was one of the key attributes in the tweets containing the correct information. These tweets were also found to be more \textit{scientific} ($-3.402$) in their content. Besides, such tweets also accounted for the presence of the elements of \textit{fear} ($-1.191$) and \textit{death} ($-1.139$). From this analysis, we see that tweets spreading misinformation tend to use more pronouns, personal pronouns (i, we, you) and certainty words (always, never). 
\section{Conclusion}
We discuss the spread of misinformation about diseases and health issues, specifically about cancer in social media (Twitter). To this end, we present a dataset of tweets with partial annotations (to be publicly released). We build classifiers to distinguish a medically relevant tweet from non-relevant ones and design sophisticated neural approaches to identify objects/item/techniques which are said to cause or cure cancer in a medically relevant tweet. We also present insights based on social aspects which are important to understand and curb the spread of misinformation. For instance, our results showed incorrect journalism is a major cause behind the spread of such misinformation. We believe this will motivate further work and research in designing methods to curb the spread of misinformation in social media.
\section{Acknowledgement}
We would like to thank Prof Mainack Mondal (mainack@cse.iitkgp.ac.in) for his continued support throughout the development of this work.

\bibliographystyle{aaai}\bibliography{bib_icwsm}

\end{document}